\documentclass[A4,12pt]{article}
\newcommand{\av}{\overline}
\newcommand{\om}{\omega}
\newcommand{\di}{\displaystyle}

\begin{document}
\begin{center}
{\Large {\bf THERMODYNAMICS OF COHERENT STATES AND BLACK HOLE
ENTROPY}}\\
\large {\bf A.G.Bashkirov}\footnote{Corresponding author.

{\it E-mail address:} abas@idg.chph.ras.ru (A.G.Bashkirov)}\\
Institute for Dynamics of Geospheres RAS, Leninskii prosp. 38
(bldg.6), 117334, Moscow, Russia\\ \large {\bf A.D.Sukhanov}\\
Peoples Friendship University of Russia, ul. Miklukho-Maklaya 8,
117198, Moscow, Russia\\


\end{center}
\begin{abstract}

Entropy and temperature of a system in a coherent state are
naturally defined on a base of a density matrix of the system. As
an example, entropy and temperature are evaluated for coherent
states of a harmonic oscillator and quantum field described by the
Klein--Gordon--Fock equation with a source term. It is shown, in
particular, that the temperature of the coherent oscillator in a
ground state coincides with the effective temperature of a
harmonic oscillator being in contact with a heat bath (Bloch
formula) when the bath temperature tends to zero. The
Bekenstein--Hawking entropy of a black hole can also be
interpreted as an entropy of coherent states of a physical vacuum
in the vicinity of a horizon surface.
\\


Klein--Gordon--Fock field; Bekenstein--Hawking entropy.
\end{abstract}

\section{Introduction}

Considerable recent attention has been focused on coherent states
in the context of a new non--Copenhagen interpretation of quantum
mechanics in which the crucial point is an interaction of the
system with its surrounding. During the interaction process the
most of states of the open quantum system becomes unstable, and,
as a result, an arbitrary state of this system becomes a
superposition of "selected states" that provides classical
description. This phenomenon has come to be known as a
decoherence.

An intrinsic connection between the decoherence phenomena and
coherent states of a system was suggested in [1]. Moreover, it was
noted [1] that decoherence might produce coherent states. This was
proved for particular case of a system of harmonic oscillators in
[2]. Analyzing a master equation for an open quantum system
interacting with a thermostat Paz and Zurek [3] found that the
coherent states were sampled as "selected states", when
self--energy of the system was of the same order as an energy of
its interaction with the thermostat.

This process is accompanied by the increase of both the open system entropy
and the thermostat entropy due to losing information about their states.
If a total closed system was initially in a pure state (with the zero entropy)
then the theorem states [4--6] that the entropies produced in the open
system and thermostat are equal.

At first glance there is a paradox of non-zero entropy and
temperature of a coherent state which is a pure state. In our
opinion, it is resolved by the fact that the coherent state is an
eigenstate of non-Hermitian operator and hence values of
observable variables remain indetermined in the coherent state,
among them an energy. Thus, it is found to be fruitful to consider
a density matrix of the coherent state and determine entropy and
temperature with the use of standard methods of statistical
physics.

Such a concept will be illustrated below (in Sec.2) by the
examples of well--known coherent states of open quantum systems,
namely, a quantum harmonic oscillator and quantum scalar field in
a vicinity of a static source.

This approach provides also fresh insight (in Sec.3) into the problem of the
black hole entropy.

\section{Entropy of coherent states}

The known isomorphism of Hilbert spaces of states of arbitrary quantum
systems ensures that a space of states of any $n$-dimensional quantum
system can be mapped on the space of states of $n$-dimensional harmonic
oscillator. Owing to this fact, properties of particular model systems
discussed below may be considered as sufficiently general.

For purposes of clarity we begin with the simplest system, that is
the one-dimensional quantum harmonic oscillator. In our approach
the oscillator in a coherent state is treated as an open system in
a steady interaction with an external force center (e.g. atomic
nucleus). The interaction with internal variables of the force
center could ensure a modification of the Shr\"odinger equation
such that it describes the decoherence of a harmonic oscillator
and the passage of its state into a coherent state. Here, however
we will confine our discussion to a prepared coherent oscillator
state.

The Hamiltonian of the system is
\begin{equation}
H=\frac 1{2m}(p^2+m^2\omega^2 q^2),
\end{equation}
where $p,\,\,q$ are respective momentum and coordinate operators
satisfying the commutation relation $[q,p]=i\hbar$. Coherent
states are conveniently [7] discussed in terms of non--Hermitian
creation and annihilation operators
\begin{equation}
a=\frac{m\om q + ip}{(2m \hbar\om )^{1/2}},\,\,\,
a^{\dag} =\frac{m \om q - ip}{(2m \hbar\om )^{1/2}} ,
\end{equation}
satisfying the commutation relations
$[a,\,a]=[a^{\dag},\,a^{\dag}]=0,\,\,[a,\,a^{\dag}]=1$.
A coherent state can be defined as an eigenstate of the
non--Hermitian operator $a$:
\begin{equation}
a|d\rangle=\left(\frac {m\om}{2\hbar}\right)^{1/2}d |d\rangle .
\end{equation}
where the eigenvalue $d$ is a complex number.

The density matrix of the $d$-coherent state is $\rho^d=
|d\rangle\langle d|$. To construct a thermodynamics of this state
we need in a partition function of $\rho^d$. Since it is defined
as a value ensuring the condition ${\rm Sp}[\rho^d]=1$, we
consider only diagonal elements of $\rho^d$. Inasmuch as coherent
states are not mutually orthogonal, we get in a coherent state
representation
\begin{equation}
\rho^d_{ff}=|\langle f|d\rangle|^2  = \exp\Big\{-\frac {m\om
|d|^2}{2\hbar}\Big\}\exp\Big\{\frac {m\om }{2\hbar}\big(-|f|^2 +
d^* f + f^* d\big)\Big\}.
\end{equation}
The first multiplier doesn't depend on the index $f$ and thus it
can be considered as a normalizing factor, that is, a reverse
value of the partition function
\begin{equation}
Q_{osc}^{d}=e^{\av {n}_d}.
\end{equation}
where $\av {n}_d= m\om |d|^2/(2\hbar)$ is the mean number of
quanta in the state $d$.

Considering that an energy representation is the most used in
statistical mechanics, we demonstrate the same result in this
representation:
$$\rho^d_{nn}=|\langle n|d\rangle|^2  = \frac 1{n!}|\langle
0|a^n|d\rangle|^2 = \frac 1{n!}\av {n}_d^n|\langle 0|d\rangle|^2 ,
$$
where the equation (3) and property $|n \rangle=
(a^{\dag})^n|0\rangle /\sqrt{n!}$ are used.  The partition
function $Q_{osc}^{d}=|\langle 0|d\rangle|^{-2}$ is determined
with the use of a condition of completeness of the set $|n\rangle
$, that is ${\rm Sp}[\rho^d]=\sum_n |\langle n|d\rangle|^2= 1$,
whence we get (5).

For the mean energy of the coherent $|d\rangle $-state of the
oscillator we have
\begin{equation}
E_{osc}^{d}= \langle d|H|d\rangle=\sum_n \rho^d_{nn} \langle
n|H|n\rangle= \hbar \omega(\av {n}_d+\frac{1}{2}).
\end{equation}

According to one of basic principles of the statistical mechanics,
all thermodynamic properties of a system can be found when the
partition function $Q$ is known. Really, in general, a
characteristic thermodynamic function $\textsf F$ is defined as
\begin{equation}
{\textsf F }= -k_B T \ln Q
\end{equation}
and known as the Helmholtz free energy $F(T,V,N)$ for a canonical
ensemble, or thermodynamic potential $J(T,V,\mu$) for grand
canonical ensemble, or Gibbs free energy $G(T,p,N)$ for $T-p$
ensemble. A thermodynamic entropy is defined as
\begin{equation}
{\textsf S}=-\frac {\partial \textsf F}{\partial T}.
\end{equation}
The energy ${\textsf E}$ is connected with the free energy
$\textsf F$ via the Gibbs--Helmholtz relation
\begin{equation}
{\textsf E}={\textsf F }+ T {\textsf S}.
\end{equation}
Thermodynamics of the $|d\rangle $-state of the oscillator is
constructed in the same manner on the base of the partition
function $Q_{osc}^{d}$:
\begin{equation}
F_{osc}^{d}=-k_B T_{osc} \av {n}_d,\,\,\,\,S_{osc}^{d}= k_B\av
{n}_d +k_BT_{osc}\frac{\partial\av {n}_d}{\partial T_{osc}},
\end{equation}
\begin{equation}
E_{osc}^{d}= k_B T_{osc}^{2} \frac{\partial\av {n}_d}{\partial
T_{osc}}.
\end{equation}
Equations (6) and (11) are used here for a self-consistent
definition of the effective temperature of the coherent state, in
contrast to common thermodynamics where the temperature is the
known parameter of a system state.

From the equations (6) and (11) we get
\begin{equation}
\hbar \omega(\av {n}_d + \frac 1{2})=k_B T_{osc}^{2}
\frac{\partial\av {n}_d}{\partial T_{osc}}.
\end{equation}
Representing $\omega$ as  $\omega= \alpha \av {n}_d$, where
$\alpha= 2\hbar /(m|d|^2)$, we can consider(12) as a differential
equation for $T_{osc}$, that is
\begin{equation}
\hbar \alpha\av {n}_d(\av {n}_d + \frac 1{2})\frac{\partial
T_{osc}}{\partial\av {n}_d }=k_B T_{osc}^{2}.
\end{equation}
Its solution is
\begin{equation}
T_{osc}=\frac{\hbar\omega}{2k_B\av {n}_d \ln(1+\frac{1}{2 \av
{n}_d} )} , \,\,\,\,T_{osc}|_{\av {n}_d \gg 1/2}= \frac{\hbar\om
}{k_B}.
\end{equation}
We can introduce an area $A_d=\pi |d|^2$ of a phase portrait
(circle) of the corresponding classical oscillator in
$(q,\,p/\omega)$ - phase plane, since $\langle q(t)\rangle _d=
d\cos \omega t$ and $\langle p(t)\rangle _d = -d\omega\sin \omega
t$. Then, $\av n_d$ and the entropy  of the coherent state are
found as
\begin{equation}
\av n_d =\frac{k_BT_{osc}m A_d}{2\pi\hbar^2} = \frac{A_d}{2\pi
l_0^2}\,\,\,\,\, {\rm and}\,\,\,\,\, S_{osc}^{d}=2k_B\av n_d
=k_B\frac{A_d}{\pi l_0^2},
\end{equation}
where $l_0=\sqrt {\hbar/(m \om)}$ is the amplitude of zero-point
oscillations.

Thus, the entropy and temperature do not vanish even for a
dynamical quantum system in the coherent state, because the
weights of eigenstates of the Hamiltonian $H$ are not
deterministic in the state.

In the opposite limit case  $\av {n} =0$ we have
$E_{osc}^0=\hbar\om/2$ and $F_{osc}^0=0$. In this case an
effective temperature $T_{osc}^{0}$ of zero-point oscillations can
be found with the use of the well-known Bloch formula (see, e.g.
[8]) for a probability distribution $b(q)$ for a harmonic
oscillator interacting with a heat bath of a temperature $T_{hb}$:
\begin{equation}
b(q)=\sqrt {\frac {m\om^2}{2\pi k_B T_{Bl}}}\,\exp \left\{-\frac
{m\om^2 q^2}{2k_B T_{Bl}}\right\},\,\,\,\, T_{Bl} = \frac{\hbar\om
}{2k_B}\coth \frac{\hbar\om }{2k_B T_{hb}}.
\end{equation}
At very low temperature of the heat bath, $k_B T_{hb} \ll \hbar\om
$, we get $T_{Bl}=T_{osc}^{0} \equiv \hbar\om /(2k_B).$ The
temperature $T_{osc}^{0}$ does not depend on $ T_{hb}$ and
determined by zero-point fluctuations entirely. The corresponding
entropy is determined with the use of the first of eq.(9) as
$S_{osc}^{0}=k_B$.

We now consider another example of an open quantum system, the
scalar field $\phi ({\bf r},t)$ described by the
Klein--Gordon--Fock (KGF) equation with a static source term in
the right hand side,
\begin{equation}
(\frac 1{c^2}\frac{\partial^2}{\partial t^2}-\nabla^2+
\frac{m^2 c^2}{\hbar^2})\phi ({\bf r},t)
=g\varrho ({\bf r}) .
\end{equation}
The density $\varrho ({\bf r})$ and intensity $g$ of the source
are given and fixed, which is natural if we treat (17) as a pure
field problem and consider only processes outside the source. On
the other hand, such a problem setting is an idealization because
we disregard all processes that occur inside the source which is a
real extended physical system. As in the previous example, we
restrict the discussion by the supposing that these internal
processes provide formatting of a coherent state of the field and
will consider the field $\phi ({\bf r},t)$ in the resulted
coherent state.

The equation (17) has been analyzed in detail in quantum field
theory (see, e.g. [9]). In particular it was shown that with
non--Hermitian operators $a_{\bf k}$ and $a^{\dag}_{\bf k}$
similar to operators (2), coherent states of the physical vacuum
are eigenstates of the operators $a_{\bf k}$.

The well--known solution for the Klein--Gordon--Fock equation with
a static source term determines the probability to find a given
number of quanta in the vacuum disregarding their moments. This
probability is governed by the Poisson distribution
\begin{equation}
\rho_{nn}=Q_{KGF}^{-1}\frac{\av {n}^n}{n!} ,\,\,\,\,\,\,n =\sum_i
n_{{\bf k}_i}.
\end{equation}
Hence we led to the same relations which were introduced above for
the particular case of the harmonic oscillator. The partition
function and entropy of this ensemble are
\begin{equation}
Q_{KGF}=\sum_{n=0}^{\infty}\frac{\bar n^{\di n}}{n!}=e^{\di \bar n
},\,\,\,\,F_{KGF}=-k_BT_{KGF}\bar n,\,\,\,\, S_{KGF}=2k_B \bar n.
\end{equation}
An energy of virtual quanta is presented as the sum of energies of
harmonic oscillators of the frequencies $\om _i$ and mean
occupation numbers $\bar n_{{\bf k}_i}$:
\begin{equation}
E_{KGF}= \av {n}\hbar \av{\om} + \frac 1{2} \hbar\av{\om} ,
\end{equation}
where $\av\om =\sum_i\om_i \bar n_{{\bf k}_i}/\bar n$ is the mean
frequency. Then, we find the thermodynamic temperature of the
field for the limit case $\bar n\gg 1/2$ as
\begin{equation}
T_{KGF}= \frac{\hbar\av\om }{k_B}.
\end{equation}
Here, as in the previous example, the coherent state temperature
is determined by the system properties and can not be decreased
without destroying the coherent state.

To estimate $\bar n$, we consider the expectation value of the
field potential of a spherical source of radius $d$, which can be
represented [9] in the ground state as
\begin{equation}
\langle 0|\phi ({\bf r})|0\rangle_{\mid_{r>d}}=g\int\,d^3 r'\,
\frac{e^{-|\bf r -\bf r'|/\lambda_c}}{|\bf r -\bf r'|} \varrho
(r') ,
\end{equation}
where $\lambda_C=\hbar/(mc)$ is the Compton radius.

Equation (22) implies that a cloud of virtual quanta envelops the
source by a spherical layer of thickness $\lambda_C $, which do
not depend on a size of the source. Therefore, $\bar n$ must be
proportional to the volume of the layer $\lambda_C\,A_d $ (here
$A_d=4\pi d^2 $ is the area of the layer surface and it is
supposed that $d\gg\lambda_C $), divided by the quantum volume
element $\sim \lambda_C^3 $:
\begin{equation}
\bar n \sim \frac {A_d}{\lambda_C^2} .
\end{equation}
Entropy (21) then becomes
\begin{equation}
S_{KGF} \sim  k_B \frac {A_d}{\lambda_C^2} .
\end{equation}
An important fact is that entropy (24) is proportional to the
source surface area (cf. the equation (15) for the harmonic
oscillator). We have such a dependence because the entropy results
from stochastic processes occurring in the vacuum near the surface
of contact with the source. The contribution from the source
itself into the entropy of the total system is not taken into
account in accordance with the initial statement of the problem.

\section{Black hole entropy}

The entropy $S_{BH}$ of a black hole is
\begin{equation}
S_{BH}=k_B\frac {A_{BH}}{4 \lambda_P^2} ,
\end{equation}
where $A_{BH}=16\pi G^2M_{BH}^2/c^4$ is the horizon area,
$\lambda_P=(\hbar \,G/c^3)^{1/2}$ Planck length and $M_{BH}$ black
hole mass. The entropy $S_{BH}$ in the form (25) was firstly found
by Bekenstein [10,11] and Hawking [12] using purely thermodynamic
arguments based on first and second laws of thermodynamics.

It is evident that the entropy (25) of a black hole $S_{BH}$ is of
the same form as the equation (24). The proportionality of the
entropy to the surface area of a black hole seemed paradoxical
over a long time. This paradox was mostly resolved in [13,14],
where the contribution to $S_{BH}$ coming only from virtual
quantum modes which propagate in the immediate vicinity of the
horizon surface was calculated. However, an ambiguity remains in
the procedure for selecting such modes and we therefore think that
the problem of justifying the expression (25) from the standpoint
of statistical mechanics is still open, which results in an
uninterrupted flow of papers on the subject.

In this respect the recent review paper by Bekenstein [15] is
worth mentioning, where a model quantization of the horizon area
was proposed in the section with the ambitious title "Demystifying
black hole's entropy proportionality to area". There the horizon
is formed by patches of equal area $\alpha \lambda_P^2$ which get
added one at a time. Since the patches are all equivalent, each
will have the same number of quantum states, say, $\kappa$.
Therefore, the total number of quantum states of the horizon is
\begin{equation}
Q_{BH}=\kappa^{\di A_{BH}/(\alpha \lambda_P^2)} .
\end{equation}
In essence Bekenstein considers his model construction as a
microcanonical ensemble for the patches. As a result, he treats
$Q_{BH}$ as the thermodynamic weight of the system and defines the
entropy of the horizon as the statistical (Boltzmann's) entropy
for the microcanonical ensemble
\begin{equation}
S^{B}_{BH}=k_B\ln Q_{BH}.
\end{equation}
To ensure that this equation would result in the desired
thermodynamic expression (25) Bekenstein puts
\begin{equation}
\alpha=\beta\ln \kappa
\end{equation}
and chooses $\beta = 4$.

It should be noted here that substituting equation (28) for
$\alpha $ into the starting equation (26) immediately results
(because $\di \kappa^{\di 1/\ln \kappa}=\di e$) in the familiar to
us form (see (5), (19) ) of the partition function of the coherent
state
\begin{equation}
Q_{BH}=e^{\av n},\,\,\,\, {\av n}= A_{BH}/(\beta \lambda_P^2) ,
\end{equation}
The Boltzmann's formula (27) for black hole entropy follows from
here if the thermodynamic definition of the entropy (8) is used
under condition that ${\av n}$ does not depend on $T_{BH}$ and the
partition function is regarded as the thermodynamic weight and
$\beta =4$. Alternatively, if $\av n \propto T_{BH}$ we are to
depart from (27) and put $\beta =8$ to ensure arriving to the
equation (25) from the definition (8).

Thus, we see that the partition function, mean number of quanta
and entropy of a black hole are of the same forms as the relevant
values for coherent states of the harmonic oscillator or quantum
field in a vicinity of the static source.

On the other hand it was found [16,17] that a strong gravitational
field provides a decoherence of a system placed in the domain of
this field. It is therefore quite natural to expect that a
coherent state of virtual excitations is formed in a vicinity of
the black hole and, as a result, the black hole's entropy is the
thermodynamic entropy of this coherent state. On behalf of such
assumption we can point to great unsolved problems in deriving the
thermodynamic black hole's entropy (25) with the use of the
standard Gibbs' equilibrium statistical ensemble and the von
Neumann formula for the entropy. Some years ago one could even
read the assertion: "...it has been shown that the
Bekenstein-Hawking entropy does not coincide with the
statistical-mechanical entropy $S^{SM}=-{\rm Tr} (\rho \ln \rho)$
of a black hole" [18]. This point of view gained acceptance in
recent years. In support of this assertion we present yet another
quotation: "there are strong hints from black hole thermodynamics
that even our present understanding of the meaning of the
"ordinary entropy" of matter is inadequate" [19].\\

It is a pleasure to acknowledge fruitful discussions with our
associates Dr. A. Vityazev.
\subsection*{References} \noindent[1] Zurek W.H., Habib S., Paz
J.P. Phys.Rev.Lett. 70, (1993) 1187--1190.\\[0pt] [2] Tegmark M.,
Shapiro H.S. Phys.Rev. E50, (1994) 2538--2547.\\[0pt] [3] Paz
J.P., Zurek W.H. Quantum limit of decoherence: environment induces
superselection

of energy eigenstates. qu--ph/9811026 (1998).\\[0pt] [4] Page D.N.
Phys.Rev.Lett. 71, (1993) 1291--1294.\\[0pt] [5] Elze H.-T., Open
quantum systems, entropy and chaos. qu--ph/9710063 (1997).\\[0pt]
[6] Kay B.S., Entropy defined, entropy encrease and decoherence
understood, and some

black--hole puzzles solved. hep--th/9802172 (1998).\\[0pt] [7]
Messiah A. Quantum Mechanics, vol.1, North-Holland Publ. Comp.,
Amsterdam, 1961.\\[0pt] [8] Landau L.D., Lifshitz E.M. Statistical
Physics, Pergamon Press, London, 1958.\\[0pt] [9] Henley E.M.,
Thirring W. Elementary Quantum Field Theory, McGraw--Hill,
N.Y.,1962.\\[0pt] [10] Bekenstein J.D. Lett.Nuovo Cimento 4,
(1972) 737.\\[0pt] [11] Bekenstein J.D. Phys.Rev. D7, (1973)
2333.\\[0pt] [12] Hawking S.W. Commun. Math. Phys., 43, (1975) 199
.\\[0pt] [13] Frolov V., Novikov I. Phys.Rev. D 48, (1993) 4545
.\\[0pt] [14] Frolov V. Phys.Rev.Lett. 74, (1995) 3319.\\[0pt]
[15] Bekenstein J.D. Quantum black holes as atoms. gr--qc/9710076
(1997).\\[0pt] [16] Hawking S.W. Commun. Math. Phys., 87, (1982)
395 .\\[0pt] [17] Ellis J., Mohanty S., Nanopoulos D.V. Phys.Lett.
B, v.221, (1989) 113--117.\\[0pt] [18] Frolov V. hep--th/9412211
(1994).\\[0pt] [19]  Wald R.M. Class.Quant.Grav. 16, (1999) A177;
gr-qc/9901033 \\[0pt]
\end{document}